%% file: twa27.tex
\documentclass[longauth]{aa}  
\usepackage{graphicx}
\usepackage{txfonts}
\usepackage{xcolor}
\usepackage{ulem}
\usepackage{url}
\usepackage{longtable}
\usepackage{multirow}
\usepackage{siunitx}
\usepackage{booktabs}
\usepackage{amsmath}
\usepackage{adjustbox}
\usepackage[colorlinks=true,allcolors=blue]{hyperref}
\usepackage{orcidlink}
\usepackage{graphicx}
\usepackage{txfonts}
\usepackage{mathtools}

\def\MJ{M_\mathrm{J}}

\def\RJ{R_\mathrm{J}}

\newcommand{\mum}{$\mu$m\xspace}

\def\c2h2{C$_\mathrm{2}$H$_\mathrm{2}$\xspace}
\def\c6h6{C$_\mathrm{6}$H$_\mathrm{6}$\xspace}
\def\c2h6{C$_\mathrm{2}$H$_\mathrm{6}$\xspace}
\def\co2{CO$_\mathrm{2}$\xspace}

\begin{document} 
% \title{Mid-infrared observations of the young planetary system TWA~27 with JWST/MIRI}
\title{JWST/MIRI observations of the young TWA~27 system: hydrocarbon disk chemistry, silicate clouds, evidence for a CPD}
%\title{The young planetary system TWA~27 through JWST/MIRI: hydrocarbon disk chemistry, silicate clouds, and evidence for a CPD}

% \input{authors}

\date{Received 7 July 2025 / Accepted xxx}

% \institute{Institute of Particle Physics and Astrophysics, ETH Zurich, Wolfgang-Pauli-Str. 27, 8093 Zurich, Switzerland}

\abstract
   {The Mid-Infrared Instrument (MIRI) on board the James Webb Space Telescope (JWST) enables the characterisation of young self-luminous gas giants in a wavelength range previously inaccessible, revealing insights into physical processes of the gas, dust, and clouds.} 
   {We aim to characterise the young planetary system TWA~27 (2M1207) in the mid-infrared (MIR), revealing the atmosphere and disk spectra of the M9 brown dwarf TWA~27A and its L6 planetary mass companion TWA~27b.}
   {We obtain data from the MIRI Medium Resolution Spectrometer (MRS) from 4.9 to 20 \mum, and MIRI Imaging in the F1000W and F1500W filters. We apply high-contrast-imaging data processing methods in order to extract the companion spectral energy distribution up to 15~\mum at a separation of 0$\farcs$78 and contrast of 60. Using published spectra from JWST/NIRSpec, we analyse the 1-20 \mum spectra with self-consistent atmosphere grids, and the molecular disk emission from TWA~27A with 0D slab models.}
   {We find that the atmosphere of TWA~27A is well fitted with the \texttt{BT-SETTL} model of effective temperature $T_{\text{eff}}\sim 2780$~K, $\mathrm{log} g \sim 4.3$, and a blackbody component of $\sim$740~K for the circumstellar disk inner rim. The disk consists of at least 11 organic molecules, and neither water or silicate dust emission are detected.
   The atmosphere of the planet TWA~27b matches with a $T_{\text{eff}}\sim$1400~K low gravity model when adding extinction, with the \texttt{ExoREM} grid fitting the best. MIRI spectra and photometry for TWA~27b reveal a silicate cloud absorption feature between 8-10 \mum, and evidence (>5$\sigma$) for infrared excess at 15 \mum that is consistent with predictions from circumplanetary disk emission.}
   {The MIRI observations present novel insights into the young planetary system TWA~27, showing a diversity of features that can be studied to understand the formation and evolution of circumplanetary disks and young dusty atmospheres.}

   \titlerunning{Observation of TWA~27 with JWST/MIRI}

    \authorrunning{P. Patapis, et al.} %

    \input{authors}
    \input{affiliations}

   \keywords{Planetary systems,
   Planets and satellites: atmospheres,
   Protoplanetary disks,
   Infrared: planetary systems,
   Methods: data analysis,
   Techniques: imaging spectroscopy
               }
   \maketitle
%
%-------------------------------------------------------------------

\section{Introduction}

% motivation
Studying young planetary systems with direct imaging provides empirical insights into the processes of planet formation and evolution \citep{Bowler2016}. The properties of their early atmospheres, the circumplanetary disk (CPD), location and interaction with the circumstellar disk determine initial conditions and bulk composition that are crucial for interpreting the atmospheres of evolved planets, the diversity of exoplanets, and even provide insights into the formation of our own solar system \citep{Oberg2011, Cridland2016}. 

Young planets have predominantly been detected and characterized in the near-infrared (NIR, 1-4 \mum), where the emission of 1000-2000 K effective temperature peaks. With the advent of the James Webb Space Telescope \citep[JWST, ][]{Gardner2023} and access to mid-infrared spectra (MIR, 4-28 \mum) with the Near Infrared Spectrograph \citep[NIRSpec, ][]{Jakobsen2022} and the Mid Infrared Instrument \citep[MIRI, ][]{Wright2015_MIRI, Wright2023_MIRI}, the properties of clouds, disequilibrium chemistry, and CPDs around gas giant exoplanet companions can be investigated \citep{Miles2023, Boccaletti2024, Worthen2024, Hoch2024_bd, Cugno2024, Hoch2025}. 
Clouds, in particular silicate condensates \citep[eg. forsterite Mg$_\mathrm{2}$SiO$_\mathrm{4}$, enstatite MgSiO$_\mathrm{3}$, quartz SiO$_\mathrm{2}$;][and references therein]{Morley2024, Moran2024}, convection \citep{Tremblin2015, Tremblin2020}, and carbon disequilibrium chemistry \citep[eg., ][]{Mukherjee2022} are thought to be the driving processes that make low gravity, young gas giants appear very red around the L/T spectral type transition \citep{Skemer2014}. These physical processes can bias the measurement of the atmospheric bulk composition, which is often used to infer formation pathways \citep{Oberg2011, Mordasini2016, Molliere2022}. For example, \cite{Miles2023} found CO at 4-5 \mum and a silicate cloud at 8-10 \mum in the spectra of the planetary mass companion (PMC) VHS 1256~b, indicating a turbulent and cloudy atmosphere. At the same time, young gas giants could still retain some of their CPD mass, which has been detected in the PDS~70 system as an infrared excess \citep{Stolker2020pds70, Christiaens2024, Blakely2024} or continuum emission with ALMA \citep{Benisty2021}. With JWST/MIRI, low resolution spectra of CPDs for low mass companions (GQ~Lup~B and YSES-1~b) have been observed up to $\sim$~12~\mum \citep{Cugno2024, Hoch2025}, opening up the possibility to investigate new aspects of their physics.

% disks of low mass stars and BDs
In terms of circumstellar disks, MIRI revealed new insights into the disks of low-mass stars and brown dwarfs \citep{Henning2024}. The MIR spectra show the emission of organic molecules like C$_\mathrm{2}$H$_\mathrm{2}$, C$_\mathrm{2}$H$_\mathrm{6}$, C$_\mathrm{6}$H$_\mathrm{6}$ and HCN, and other than CO$_\mathrm{2}$, a lack of oxygen-bearing species typically seen in protoplanetary disks around massive stars like H$_\mathrm{2}$O \citep{Tabone2023, Arabhavi2024}. Furthermore, these disks rarely show any emission from silicate dust around 10 \mum, implying that the dust has settled on the midplane of the disk and has grown to larger particle sizes \citep{Tabone2023}. These observations and upcoming programs from JWST provide valuable laboratories for studying the physical properties of disks down to the planetary mass regime, which could be in turn used to interpret the spectra of disks around young companions revealed through high-contrast imaging \citep{Cugno2024}.%\danny{Although we do have examples of VLM disks with silicate emission: Kanwar+ 2024, and even disks WITH clear H2O emission: Xie+ 2023 (on Sz 114).} \ama{Just a note on Sz114: it is a borderline case of VLMS and LMS aka TTauris, with about 0.2Msun and 0.2Lsun, and Xie+2023 never refer to it as a VLMS. The MINDS VLMS sample are distinctly lower mass and luminosity.} \ama{Morales-Calderon et al. in prep. show water+dust feature around the VLMS, [NC98] Cha Ha 1.} \ama{Arabhavi et al. in prep., the VLMS overview, also show diverse dust features in the MINDS sample.}
% atmospheres of young gas giants
% young systems betaPic, PDS70, GQLup with JWST
% 2M1207/TWA~27 

The young planetary system TWA~27 (2MASSW J1207334-393254, also known as 2M1207) is part of the TW~Hya stellar association with an estimated age of 10 Myr \citep{Hoff1998}, at a distance of 64.5$\pm$0.4 pc \citep{Bailer-Jones2021}. 
The primary, TWA~27A, is a late M type (M9) $\sim$25~$\MJ$ brown dwarf, which has a compact disk detected with Spitzer and ALMA \citep{Riaz2007, Riaz2008, Morrow2008, Ricci2017}. The companion\footnote{The first directly imaged PMC to be discovered.} orbiting at 50~AU (0$\farcs$78) is a $\sim$5~$\MJ$ PMC \citep{Chauvin2005} that, since its discovery, has been problematic to fit with atmospheric models, due to its unusual red colour and late L spectral type \citep[eg., ][]{Patience2010, Barman2011, Skemer2011, Skemer2014}. \cite{Zhou2016, Zhou2019} observed the planet with the Hubble Space Telescope (HST) and found tentative rotational modulation with a period of $\sim$11 hours in the first epoch but not the second. These modulations could be a sign of patchy  cloud coverage, as seen in time series observations of brown dwarfs \citep{Vos2023}. %Sterzik et al. 2004, Scholz et al. 2005 report variability from days to weeks, Stelzer et al. 2007 report variability from hours to years. 

JWST observed the system with the NIRSpec Integral Field Unit \citep[IFU, ][]{Boeker2023} with program identifier (PID) 1270 (PI: S. Birkmann), obtaining a full 1-5 \mum spectrum at a resolution of R$\sim$2700 \citep{Luhman2023_twa27, Manjavacas2024}. Using the JWST/NIRSpec observations, the atmosphere of TWA~27A was thoroughly studied by \cite{Manjavacas2024}. They find that the photosphere of the brown dwarf is best fit with T$_\mathrm{eff}$=2600~K, log~\textit{g}=4.0, and a radius of 2.7 $\RJ$.
\cite{Luhman2023_twa27} found that the 1-5 \mum spectrum of TWA~27b fits with a \texttt{ATMO} cloudless model \citep{Tremblin2015} of 1300~K and low gravity, noting the muted molecular features, and complete lack of methane compared to the similar spectral type but older VHS 1256~b. Furthermore, \cite{Luhman2023_twa27} found, for the first time, emission from hydrogen recombination lines (Paschen-$\alpha, -\beta, -\gamma, -\delta$), which for young objects typically imply accretion of gas from the circumplanetary environment. Subsequent analysis by \cite{Marleau2024} provided an upper limit on H$\alpha$ from HST data, and combined with the JWST/NIRSpec line luminosities, estimated a mass accretion rate $\Dot{\mathrm{M}}$ between 0.1 $\times 10^{-9}$ -- 5 $\times 10^{-9}\,\MJ~\mathrm{yr}^{-1}$. \cite{Aoyama2024} followed up, attempting to constrain the accretion properties and geometry, hinting towards magnetospheric accretion. The atmosphere of the companion was further investigated using self-consistent radiative convective models\citep{Manjavacas2024}, and free atmospheric retrievals \citep{Zhang2025}. The latter point to an atmosphere with thick, patchy clouds as already noted in previous works.

In this work we present the MIRI observations of the TWA~27 system. In Section~\ref{sec:observations} we describe the observations, data reduction, and high contrast imaging methods used to obtain the spectra of the two sources. In Section~\ref{sec:analysis} the analysis of the atmospheres with self consistent models and slab models of the disk are presented. The results are discussed in Section~\ref{sec:discussion}, and we conclude our work in Section~\ref{sec:conclusion}.

\section{Observations and Data Reduction}\label{sec:observations}

\begin{figure*}[h]
    \centering
    \includegraphics[width=\textwidth]{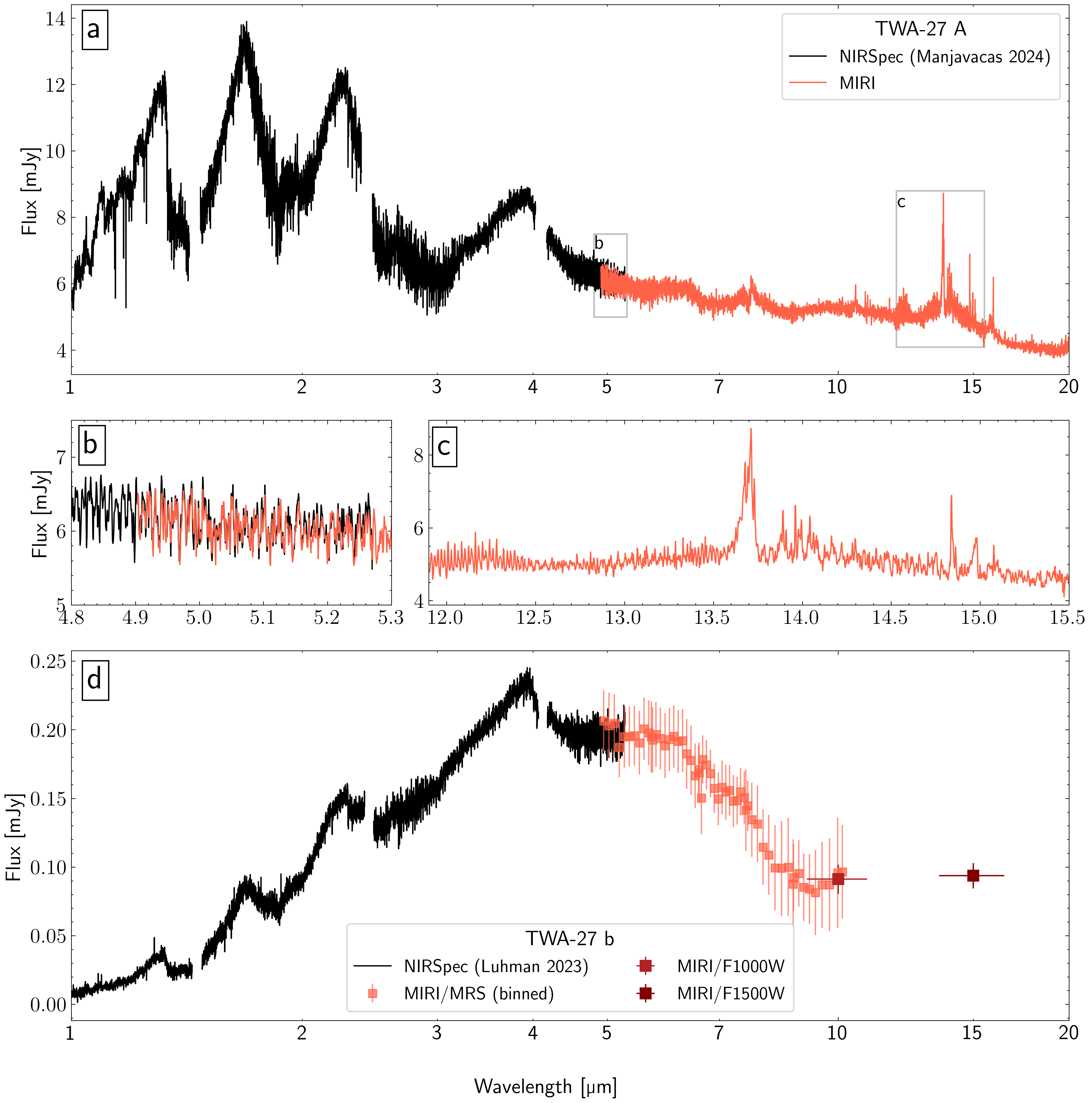}
    \caption{Complete 1-20 \mum spectra of TWA~27A (a) and TWA~27b (d) with JWST/NIRSpec and JWST/MIRI. Details of TWA~27A are presented in panel (b) for the instrument overlap between NIRSpec and MIRI, and in panel (c) for the molecular emission of the TWA~27A disk. Error bars for the NIRSpec spectra, and MIRI spectrum of TWA~27A are not visible.}
    \label{fig:twa-27-A}
\end{figure*}

\subsection{Observations}
% science
TWA~27 was observed on February 7th 2023, as part of joint NIRSpec and MIRI European Consortium Guaranteed Time Observations (GTO) under PID 1270 (PI: S. Birkmann). MIRI observed the system in two filters (10 \mum : F1000W, 15 \mum : F1500W) with the MIRI Imager \citep{Dicken2023_MIRI}, and with the Medium Resolution Spectrometer \citep[MRS, ][]{Argyriou2023_MRS}. The imaging mode was acquired using the full frame of the detector with 9 and 5 groups per integration for F1000W and F1500W respectively, and a 4 point extended source optimized dither pattern. The MRS used 76 groups per integration for each of the spectral channels (1-4) and three dichroic/grating settings (SHORT or A, MEDIUM or B, LONG or C), with a 4 point dither optimized for an extended source. The extended source dither pattern was selected in order to minimize the risk of placing the companion outside the field of view given some uncertainty in the pointing accuracy of the MRS in cycle 1. This is no longer the case as the accuracy of the MRS when using target acquisition is $\sim$30 mas \citep{Patapis2024}. No dedicated background observations or reference star observations were acquired for MIRI. In contrast, the NIRSpec observation strategy included a nearby reference star observation of a similar spectral type as TWA~27A \citep[TWA~28, ][]{Luhman2023_twa27} which greatly simplified the subtraction of the primary point spread function (PSF). 

% ref stars
In order to remove the PSF for the MIRI observations and extract the spectra and fluxes of the companion we relied on publicly available data containing stars that could be used to model the PSF, detailed in Section~\ref{sec:psf_sub}. The MRS reference stars were selected from commissioning and cycle 1 calibration programs (PIDs: 1050, 1524, 1536, 1538), with a total of 17 observations of different stars. The selection was mainly based on availability of the data and vetting of the sources as single stars since they were used as calibration stars for the MRS \citep{Gordon2022}. The Imager reference stars were selected from PID~1029 for F1000W, and PID~4487 for F1500W. Several other programs and stars were explored but most of the stars had either too much, or not enough flux to match the flux level of TWA~27A. Fainter stars would not model the wings of the PSF well enough, and introduce additional noise to the data by scaling up the noise from the reference star, whilst bright stars have a more pronounced brighter-fatter-effect \citep[BFE, ][]{Argyriou2023_bfe, Gasman2024} that changes the flux distribution in the PSF core. 

\subsection{Pipeline processing}\label{sec:pipeline}

All MIRI data were processed with the \texttt{jwst} pipeline\footnote{pipeline version 1.12.5, CRDS version 11.17, CRDS context jwst\_1197.pmap}. The MRS data were downloaded at the rate file level from the Mikulski Archive for Space Telescopes (MAST), and subsequently processed with the spectroscopic pipeline \texttt{Spec2Pipeline} using the default MRS steps except the residual fringing correction. The IFU cubes were built with the \texttt{Spec3Pipeline} using the drizzle algorithm in IFUALIGN mode, the internal coordinate system of the MRS that results in the same PSF alignment, irrespective of the given V3 position angle of the telescope during an observation \citep{Law2023}. The MIRI Imager data were directly downloaded from MAST in the final calibrated (i2d.fits) stage.

%backgrounds
We note that we treated the background subtraction differently for the extraction of the flux from the primary brown dwarf TWA~27A, and the companion TWA~27b. The MRS suffers from two types of structured noise that we consider under the general term background. The telescope thermal background and astrophysical background is generally a spatially smooth component, varying exponentially with wavelength, relevant for wavelengths longer than 15 \mum. For shorter wavelengths, vertical detector stripes (approximately along the dispersion direction on the detector) which are thought to be originating from varying dark current \citep{Argyriou2023_MRS}, present a structured background in the MRS spectral cubes. When observing faint sources these detector stripes can be a limiting factor, dominating the signal, and due to the variability of the stripes even observing a dedicated background cannot adequately remove them. %We detail the various background features and background subtracted data in \todo{Appendix~\ref{ap:bkg}}.

For the primary we made use of a nod subtraction strategy, using the large offsets between individual MRS dithers at the detector level, subtracted from each other to remove the background. This method does not work for the companion, since the dither offsets and orientation of the system cause subtraction of the science flux when applying the nod subtraction. We estimated an empirical background using the rate files of the science observations by taking the minimum value of each pixel among the four dithers, sigma clipping for outliers to produce a smooth background image. The offset (or pedestal) in flux seen in the first exposure of each observation was removed beforehand. We subtracted the estimated background from each dither, and processed the data with the \texttt{jwst} pipeline as described before. For the MIRI Imager, subtracting the median of all pixels including signal (ie. excluding the region of the Imager mask) effectively removed the background. 

%spectral extraction
For the spectral extraction of TWA~27A we used the \texttt{Spec3Pipeline} method \texttt{extract\_1d} with an aperture size of 1.5 Full Width at Half Maximum (FWHM) in order to avoid the negative PSFs from the nod-subtraction, with the residual fringe correction applied to the extracted spectrum, and no additional annulus background subtraction. Even with such a small aperture radius, the nod-subtraction impacts the extraction introducing a systematic as a function of wavelength. In order to correct this and still benefit from the smaller noise of the nod-subtraction, we reduced the data again without any background subtraction, and followed the default pipeline that calculates the background from an annulus around the source. We fitted a smooth baseline to each of the two spectra (with and without background subtraction), and corrected the difference between them by adding the flux discrepancy to the nod-subtracted spectrum.

For NIRSpec we used the publicly available, calibrated spectra of TWA~27A and TWA~27b by \cite{Luhman2023_twa27} and \cite{Manjavacas2024}. In Figure~\ref{fig:twa-27-A}a we present the 1-20 \mum spectrum of TWA~27A. In the middle panels (b, c) of Figure~\ref{fig:twa-27-A} we show two noteworthy features of the spectrum; the overlap wavelength range and flux agreement between MIRI and NIRSpec (left) and rich molecular emission from hydrocarbons (right) examined in detail in Section~\ref{sec:twa-27_A_molec_emission}.

\subsection{PSF Subtraction and Spectral Extraction of the Companion}\label{sec:psf_sub}
\begin{figure*}[h]
    \centering
    \includegraphics[width=\textwidth]{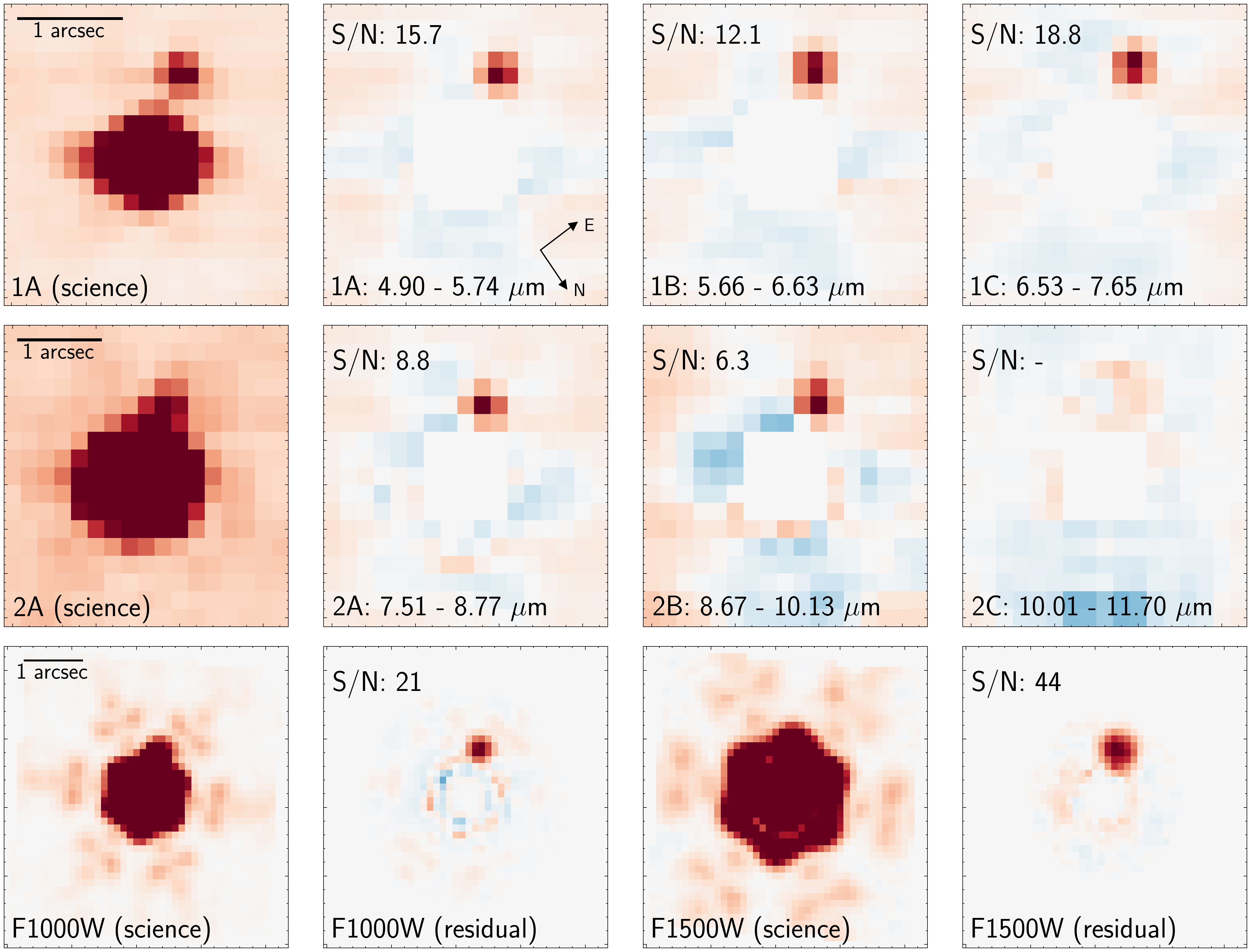}
    \caption{JWST/MIRI pipeline processed science frames of TWA~27b, and residuals after subtraction of the PSF of TWA~27A. The first two rows show the MRS median combined IFU cubes in Channels 1 and 2, while the third row shows the two MIRI Imaging filters at 10 and 15 \mum. The S/N is denoted for each frame and the scale is set to the maximum flux of each image   .}
    \label{fig:psf_sub}
\end{figure*}

\paragraph{\textbf{MRS.}} For the PSF subtraction and spectrum extraction of the MRS observations, we follow the same procedure as \cite{Cugno2024}, who looked at the high contrast companion GQ Lup B with the MRS and achieved excellent PSF subtraction. For each spectral channel of the MRS we identify the brightest pixel in the median combined (along the wavelength axis) cube, and crop each frame (with a size of 19 pixels for channel 1 and 17 pixels for channel 2). The cubes are not aligned to the centre of the PSF in order to avoid interpolation artefacts, and the centre itself is estimated by fitting a 2D Gaussian to the median combined cube. The reference stars are aligned to the precise sub-pixel position of TWA~27A using spline interpolation, and cropped to the same size as the science frames. We mask the central $0\farcs5$ and apply principal component analysis \citep[PCA, ][]{Amara2012} to the stack of reference stars in order to model the PSF. In Figure~\ref{fig:psf_sub} the median combined residuals of the PSF subtracted TWA~27 data are shown for each spectral band. Denoted as ``science'' frames are the median combined cubes prior to the PSF subtraction.  

The spectral extraction is performed by injecting a negative PSF from a well characterised standard star (PID 1536, observation 22 as in \citealt{Cugno2024}) at the position of the companion and minimising the residuals within an aperture of 1.5 FWHM at the position of the companion. This process yields a contrast between the companion and the standard star, and the companion spectrum is calculated by multiplying the measured contrast with the spectrum of the standard star. The position of the companion is fit with a 2D Gaussian on the mean combined residual cube. The uncertainty estimation is also performed in a similar manner, injecting positive PSFs of the same flux as the extracted companion at the same separation and different position angles around the primary. The flux of the injected signal is retrieved with the same method as the companion, and the uncertainty is calculated as the standard deviation between the injected and retrieved signal.

The S/N of each wavelength bin up to 8 \mum is greater than 5, and it decreases to $\sim$ 3 at 10 \mum. In band 2C and Channel 3 the companion could not be reliably extracted, while in Channel 4 it was not detected at all. We note that due to the systematic behaviour of the noise in the residuals, binning along wavelength improves the S/N non-linearly. The extracted spectrum was found to be underestimated with respect to the NIRSpec spectrum by 20\%. The flux in the wavelength overlap between the MRS bands was well within the error, pointing to a single systematic offset. The PCA projection was initially suspected to remove signal from the companion, despite this not being the case in \cite{Cugno2024}. To test this, we extracted the spectrum in band 1A of the MRS without performing the PSF subtraction since the two point sources are sufficiently separated at these wavelengths (see the top left frame in Figure~\ref{fig:psf_sub}). The extracted spectrum was still biased and practically indistinguishable from the PCA extracted spectrum. Next we tested whether the empirically estimated background described in Section~\ref{sec:pipeline} was affected the flux. Indeed, we found that extracting the spectrum in cubes processed without the background subtraction explained most of this offset. However, the detector stripes limited the quality of the spectrum, introduced artefacts, and worsened the S/N. Hence we opted for adding a single systematic offset to each band in order to match the NIRSpec flux, but avoided the detector systematics that would have dominated the spectrum otherwise. The spectrum of TWA~27b binned every 100 wavelengths is shown in Figure~\ref{fig:twa-27-A}d. %The offset was estimated by binning the NIRSpec spectrum in the overlap region in the same way and calculating the mean of the difference between NIRSpec and MRS. 

\paragraph{\textbf{Imager.}}
The PSF subtraction for the two MIRI filters was performed with a single reference star. TWA~27A was aligned to the brightest pixel, cropped (size of 41 pixels), and the central $0\farcs5$ was masked. The reference star was aligned to the sub-pixel location of the science star in the same way as with the MRS. We masked the position of the companion and the residuals in an annulus around the brown dwarf were minimised. The images before and after PSF subtraction are shown in the bottom row of Figure~\ref{fig:psf_sub} for each filter. Since the subtraction worked well and the companion is not expected to suffer from any over-subtraction as in the PCA method, we extracted the flux using aperture photometry, with the location of the companion fitted with a 2D Gaussian, an aperture of 2.8 and 3.7 pixels for F1000W and F1500W respectively, and aperture correction for the given apertures applied. The error was calculated by placing apertures at the same separation and different position angles around the primary, extracting the flux of the residual noise, and calculating the standard deviation from these noise apertures, accounting for small sample statistics \citep{Mawet2014}. The S/N was estimated at $\sim$21 for F1000W and $\sim$44 for F1500W. The two flux points are plotted in Figure~\ref{fig:twa-27-A}, and without applying any offset the 10 \mum flux agrees well with the flux measured with the MRS, thus reassuring our empirical offset correction for the MRS.

\section{Analysis}\label{sec:analysis}
\subsection{Self-consistent grid models}
The emission of the brown dwarf primary and PMC were modelled using pre-computed self-consistent atmosphere model grids. We used the \texttt{python} package \texttt{species} \citep{Stolker2020_species} that provides tools for fitting the atmosphere grids to the data using a Bayesian framework. We employed nested sampling with 2000 live points using the \texttt{PyMultinest} package \citep{Buchner2014} to fit the spectra and calculate the posterior distribution of the free parameters for each model (eg. see Table~1 in \cite{Petrus2024}). The atmosphere models, described next, are commonly applied to L/T transition objects like TWA~27b \citep{Boccaletti2024, Petrus2024}.

\textbf{ATMO.} The cloudless ATMO grid has been developed for brown dwarfs exploring the effect of disequilibrium chemistry on the temperature gradient of the atmosphere due to fingering convection \citep{Tremblin2015}. The model has been adapted to calculate evolutionary tracks for gas giants \citep{Phillips2020}, as well as dedicated grids for L/T transition objects including an adiabatic index parameter that controls the change in the temperature gradient \citep{Petrus2023}. In this work we use the latter for fitting the atmosphere.

\textbf{ExoREM.} This 1D radiative-convective equilibrium model was developed specifically for L/T transition objects, including a self consistent treatment of clouds \citep{Charnay2018}. While the model includes silicate clouds, it does not reproduce the 8-10 \mum silicate absorption seen in brown dwarfs and sub-stellar companions \citep[eg. ][]{Miles2023}.

\textbf{SONORA Diamondback.} Similar to ExoREM the Sonora Diamondback \citep{Morley2024} grid includes clouds and calculates the atmosphere in radiative-convective equilibrium. The difference is that the model is coupling the thermal evolution of these gas giants with cloudy atmospheres.

\textbf{BT-SETTL.} For TWA~27A we need to consider its higher effective temperature, which only the ATMO grid covers \citep{Phillips2020}. Therefore, we add another grid BT-SETTL\footnote{CIFIST} \citep{Allard2012} that is appropriate for the young brown dwarf and includes equilibrium chemistry and the effect of clouds.

% \section{Results}\label{sec:results}
\subsection{TWA~27A model fit}
\begin{figure*}[h]
    \centering
    \includegraphics[width=\textwidth]{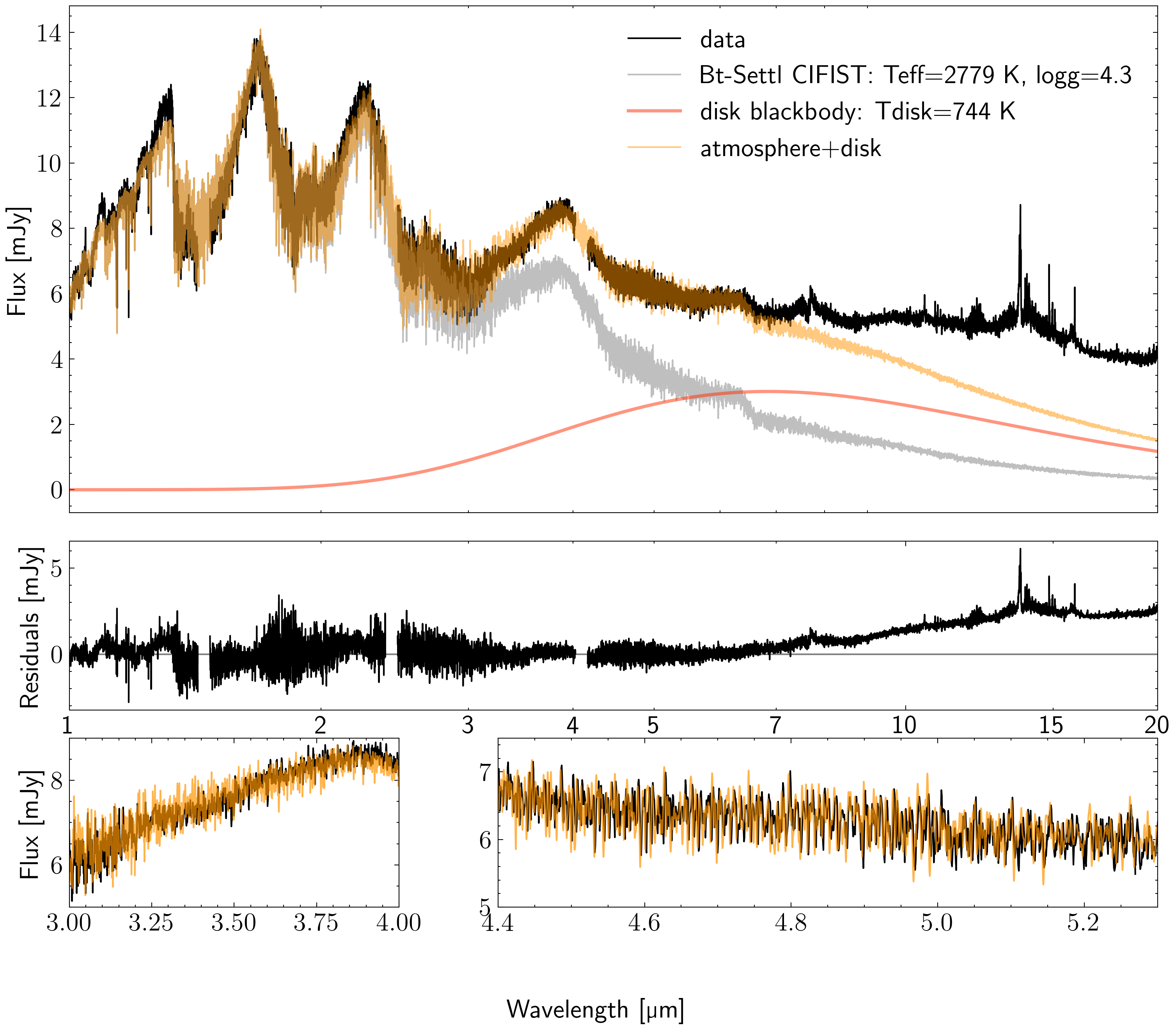}
    \caption{Model fit to TWA~27A. Top: data and model fit using the \texttt{BT-SETTL} grid. The atmosphere model is shown in gray, the blackbody disk component in red, and the combined model in gold. Middle: residuals between data and model in physical units (mJy). Bottom: details of model fit for two regions of the spectrum.}
    \label{fig:twa-27-A_modelfit}
\end{figure*}

    We tested three model setups for TWA~27A; just using \texttt{BT-Settl} as the atmosphere model, adding extinction by dust (here ISM dust, reported as the magnitude of visual extinction), and adding a blackbody component that fits the IR excess \citep{Stolker2021, Cugno2024}. As seen in Figure~\ref{fig:twa-27-A}, the disk emission dominates the mid-infrared flux, and thus we restricted all fits to include the spectrum up to 6.5 \mum where the molecular emission of the disk is not as prominent. The atmosphere model alone did not match the SED particularly well. Compared to results from \cite{Manjavacas2024} using the same model setup with \texttt{BT-SETTL} (ie. no extinction and no disk contribution), our best-fit parameters have lower temperature ($T_{\mathrm{eff}} \sim 2380$~K) and gravity ($\mathrm{log} g \sim 3.5$), with a radius of 3.3 $\RJ$. Adding extinction ($A_V \sim 1.2$~mag) increased the temperature ($T_{\mathrm{eff}} \sim 2640$~K) but the gravity remained low. Both models predict very low masses of 12-14 $M_J$, inconsistent with what would be expected for the age of the brown dwarf \citep{Venuti2019, Marley2021}. With the addition of a disk component the best-fit parameters listed in Table~\ref{tab:model_results} agree better with \cite{Manjavacas2024}, and the model matches the overall spectrum well, as seen in Figure~\ref{fig:twa-27-A_modelfit}. At shorter wavelengths (1.3-1.5 \mum) the model underestimates the flux, but for wavelengths longer than 1.8 \mum the fit appears better than in \cite{Manjavacas2024} (see Figure 10), presumably due to the additional contribution of the disk emission, shown in the lower panels of Figure~\ref{fig:twa-27-A_modelfit}. The blackbody component models a single temperature ($T_{\mathrm{BB}}\sim$744 K) that corresponds to the dust emission in the inner rim of the disk heated up by the brown dwarf.% Using Stefan-Boltzmann's law (eg. equation~1 in \cite{Cugno2024}), we calculate the radius of inner disk cavity to be 18 $\RJ$. 

\subsection{TWA~27b model fit}\label{sec:analysis_b}
\begin{figure*}[h]
    \centering
    \includegraphics[width=\textwidth]{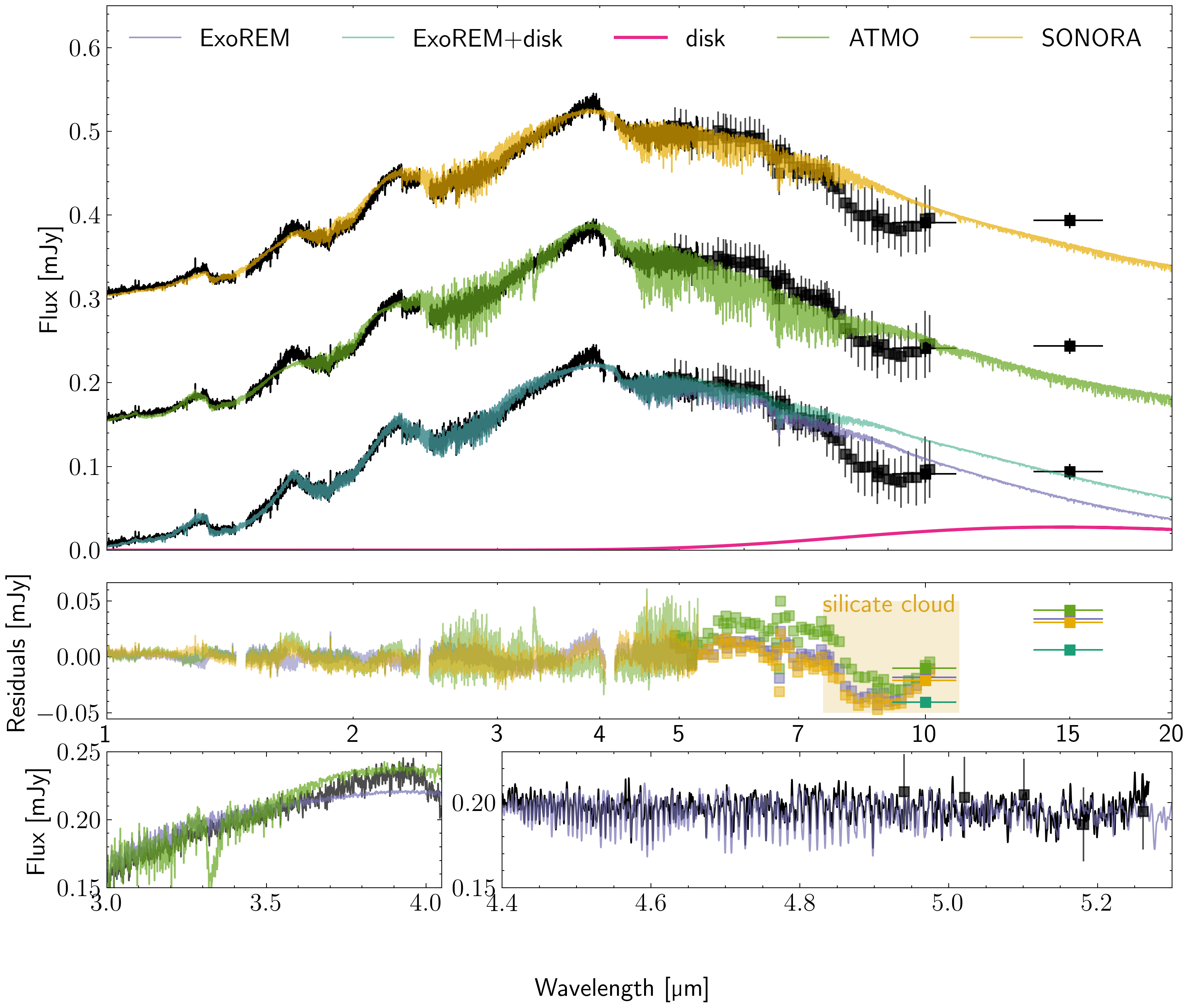}
    \caption{Model fit to TWA~27b. Top: data and model fit using the \texttt{ATMO}, \texttt{ExoREM}, and \texttt{SONORA} grids, including extinction and an offset of 0.15 mJy for visibility. In order to explain the 15 \mum IR excess due to a cold disk a blackbody component (350~K) is shown in magenta and added to the \texttt{ExoREM} flux. Middle: residuals between data and model in physical units (mJy). The discrepancy of data and models at 7.5-10 \mum is most likely attributed to a silicate cloud. Bottom: details of model fit for two regions of the spectrum, including the methane and CO bands, only showing \texttt{ExoREM}.}
    \label{fig:twa-27-b_modelfit}
\end{figure*}

\begin{table*}[]
    \centering
    \caption{Results of model fitting. Best-fit parameters of the various setups that were tested. The ``ext'' denotes a model with ISM extinction, and ``disk'' a model including an additional blackbody component.}
    \begin{tabular}{l c c c c c c c | c}
    \toprule
        Model &  T$_{\mathrm{eff}}$ [K] & log(\textit{g}) & R$_p$ [R$_J$] & C/O & Fe/H & fsed & A$_V$ [mag] & $\chi^2$\\ \hline 
        \multicolumn{8}{c |}{TWA~27A} &\\
        \texttt{BT-Settl} & 2360 & 3.5 & 3.4 & solar & solar & - & - & 204\\
        \texttt{BT-Settl+ext+disk} & 2780 & 4.3 & 2.6 & solar & solar & - & 0.7 & 41\\ \hline
        \multicolumn{8}{c |}{TWA~27b} &\\
        \texttt{ExoREM} & 1200 & 3.0 & 1.27 & 0.8 & 0.7 & - & - & 12\\
        \texttt{ExoREM+ext} & 1400 & 4.0 & 1.2 & 0.55 & -0.2 &  &3.7 & 10\\
        \texttt{ATMO} & 1280 & 3.2 & 1.15 & 0.3 & -0.2 & - & - & 26 \\
        \texttt{ATMO+ext} & 1400 & 3.2 & 1.01 & 0.4 & -0.6 & - & 2.1& 15\\
        %\texttt{SONORA} & 1250 & 4.4 & 1.3& solar & %0.5 & 1.0 & -\\
        \texttt{SONORA+ext} & 1250 & 4.4 & 1.3& solar & 0.5 & 1.0 & 2.7 & 13\\  
    \bottomrule 
    \end{tabular}
    \label{tab:model_results}
\end{table*}
Young gas giants around the L/T transition like TWA~27b have been difficult to model, showing very red colour, dusty atmospheres and muted molecular features \citep{Stolker2020pds70, Malin2024}. We followed a strategy of increasing model complexity to fit the \texttt{ExoREM, ATMO, SONORA} grids to the spectrum of TWA~27b. Here we use the full SED and apply weights to the data points\footnote{see documentation of \texttt{species}: \url{https://species.readthedocs.io/en/latest/}}, such that the MIRI photometric fluxes are equally contributing to the likelihood calculation as the higher resolution spectra. The fit results are listed in Table~\ref{tab:model_results}.

Without the addition of extinction \texttt{ATMO} does not fit the data well, especially in the 2-4 \mum region. With extinction the overall \texttt{ATMO} fit appears better, notably capturing the peak of the SED around 4 \mum, predicting higher temperature, lower gravity than the fit without extinction, and a low radius which might not be consistent with the young age of the companion. The adiabatic index parameter is constrained in both cases to 1.01, and as noted in \cite{Luhman2023_twa27} and \cite{Manjavacas2024} the model predicts a methane feature at 3.3 \mum which is not seen in the data (bottom left panel in Figure~\ref{fig:twa-27-b_modelfit}). \texttt{ExoREM} finds well matching solutions with and without extinction, with the difference being in the temperature, log(\textit{g}), and Fe/H. It fits all wavelengths better than \texttt{ATMO} except for the 4 \mum peak seen in Figure~\ref{fig:twa-27-b_modelfit}. The clouds included in \texttt{ExoREM} also suppress the methane absorption feature. The \texttt{SONORA} grid predicts a lower temperature even when extinction is applied, and estimates a higher gravity than the other models. 

None of the models reproduce the 8-10 \mum silicate cloud feature \citep[see ][]{Miles2023, Hoch2025}, evident in the residuals shown in the middle panel of Figure~\ref{fig:twa-27-b_modelfit}. Modelling this feature requires a dedicated atmospheric retrieval analysis \citep[eg. see work by ][]{Burningham2021, Vos2023} which is left for future work. The main remaining unexplained feature in the data is the slightly higher flux at 15 \mum, and thus we added a single blackbody component to model potential emission from a CPD. With only limited data in the mid-infrared, and the models not including the silicate absorption we fixed the atmosphere model to the best-fit of each grid and added the additional blackbody emission. Only using the MIRI/MRS 5-6.5 \mum and the 15 \mum MIRI/Imager fluxes, we calculated the $\chi^2$ for a grid of disk temperatures and emitting areas (T$_{\mathrm{BB}}$: 200-500 K, R$_{\mathrm{BB}}$: 1-7 $\RJ$). The parameters are quite degenerate given the available data, with temperatures of 250-450 K for decreasing emitting areas fitting the SED equally well. This range of temperatures would yield cavity sizes of 5.6-18 $\RJ$, using Stefan-Boltzman's law \citep[see Equation 1 in][]{Cugno2024}. For illustration, in Figure~\ref{fig:twa-27-b_modelfit} we plotted the \texttt{ExoREM} minimum $\chi^2$ parameter combination (T$_{\mathrm{BB}}$=350 K, R$_{\mathrm{BB}}$=2.9 $\RJ$).

% \begin{table*}[]
%     \centering
%     \caption{Results of model fitting. Best-fit parameters of the various setups that were tested. The ``ext'' denotes a model with ISM extinction, and ``disk'' a model including an additional blackbody component. R$_{\mathrm{cav}}$ is the computed size of the cavity given the luminosity of the central sourc and the estimated blackbody temperature.}
%     \begin{tabular}{l c c c c c c c c | c}
%     \toprule
%         Model &  T$_{\mathrm{eff}}$ [K] & log(g) & R$_p$ [R$_J$] & C/O & Fe/H & A$_V$ [mag] & T$_{\mathrm{BB}}$ [K] & R$_{\mathrm{cav}}$ [R$_J$] & $\chi^2$\\ \hline 
%         \multicolumn{9}{c |}{TWA~27A} &\\
%         \texttt{BT-Settl+ext+disk} & 2780 & 4.3 & 2.6 &  & & 0.7 & 744 & 18 &\\ \hline
%         \multicolumn{9}{c |}{TWA~27b} &\\
%         \texttt{ExoREM} & 1200 & 3.0 & 1.27 & 0.8 & 0.7 &  & & &\\
%         \texttt{ExoREM+ext} & 1400 & 4.0 & 1.2 & 0.55 & -0.2 & 3.7 & & &\\
%         \texttt{ExoREM+ext+disk} & 1400 & 4.0 & 1.2 & 0.55 & -0.2 & 3.7 & 250 & 18 &\\
%         \texttt{ATMO} & 1280 & 3.2 & 1.15 & 0.3 & -0.2 &  & & 35\\
%         \texttt{ATMO+ext} & 1400 & 3.2 & 1.01 & 0.4 & -0.6 & 2.1 & & & 14\\
%         \texttt{ATMO+ext+disk} & tbd & 2.7 & 1.04 & 0.6 & -0.5 & 2.6 & 250-450 & 18-5.6 &\\
%         \texttt{SONORA} & 1250 & 4.4 & 1.3& & & & & &\\
        
%     \bottomrule 
%     \end{tabular}
%     \label{tab:model_results}
% \end{table*}
\subsection{Molecular Emission from TWA~27 A disk}\label{sec:twa-27_A_molec_emission}

\begin{figure*}[h]
    \centering
    \includegraphics[width=\textwidth]{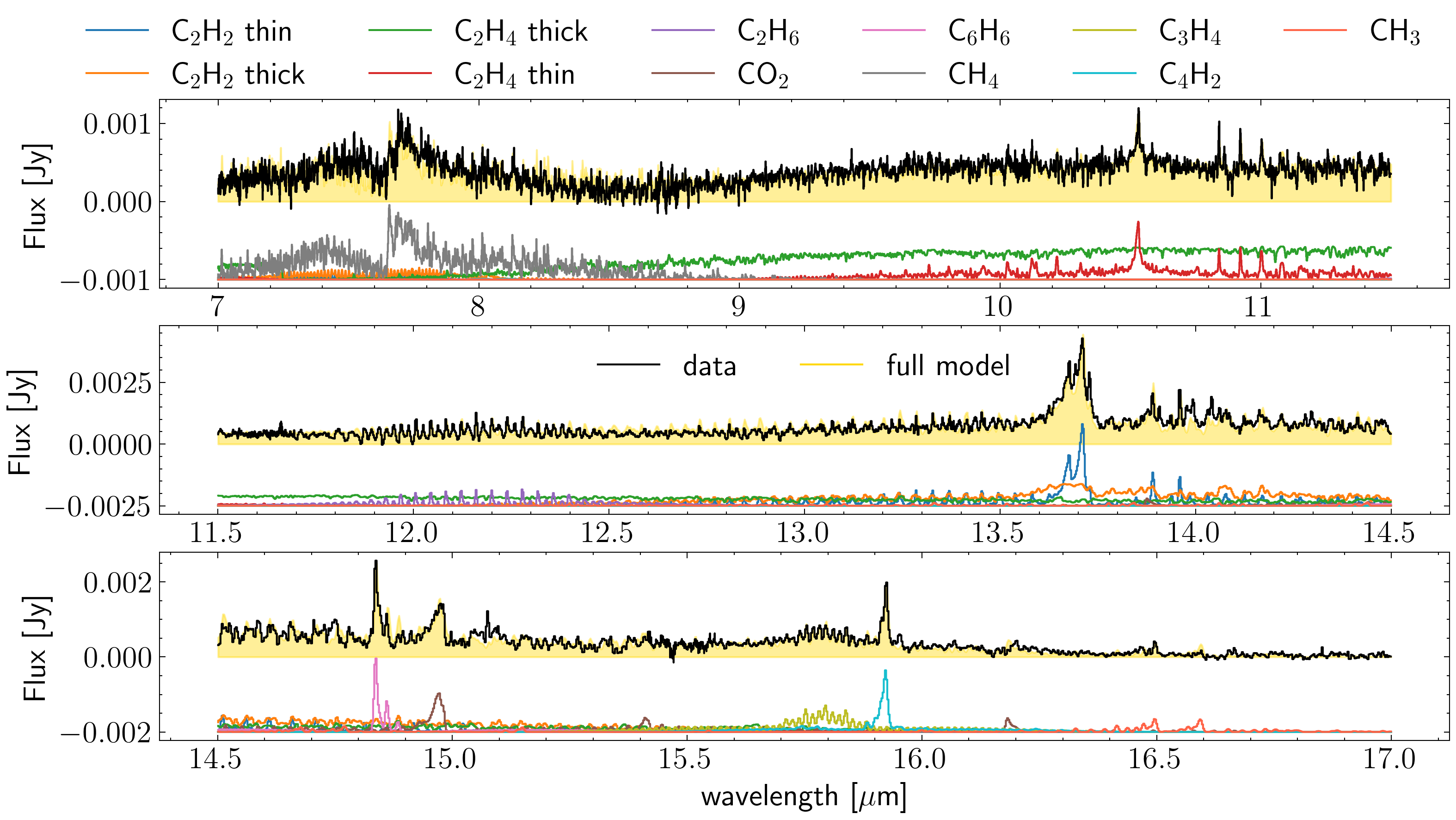}
    \caption{Disk emission of TWA~27A. Each panel covers a part of the spectral range, up to 17 \mum. The data are continuum subtracted.}
    \label{fig:twa-27-A_molfit}
\end{figure*}

As seen in Figure~\ref{fig:twa-27-A}c, the MIRI spectrum of TWA~27A reveals the presence of several molecular species. To identify the molecules and estimate their column density and temperature for each of them we used a slab modeling approach which has been extensively used to analyze emission in several other MIRI spectra (e.g. \citealt{2023ApJ...947L...6G,Tabone2023, Arabhavi2024, Kanwar2024}). In slab modeling, a small portion of the disk is approximated as a slab with uniform properties (temperature, density, velocity, and molecular composition). This simplification allows for calculations of emission without needing to model the entire complex 3D disk structure. We will follow the same procedure as \citet{Arabhavi2024}.

We use the synthetic spectra produced in \cite{Arabhavi2024} where line positions, Einstein A coefficients, statistical weights, and partition functions were taken from the HITRAN 2020 database \citep{Gordon2022hitran} (while, GEISA database -\citealt{Delahaye2021} and \citealt{Arabhavi2024} for C$_3$H$_4$ and C$_6$H$_6$).
These grids of spectra were computed varying the column density from 10$^{14}$ to 10$^{24.5}$\,cm$^{-2}$, in steps of 1/6 dex and temperature from 100 to 1500\,K in steps of 25\,K.
For C$_6$H$_6$ and C$_3$H$_4$ the temperature is limited to 600 K. Carbon isotopologues are also included in our slab model analysis as summarized in \citet{Arabhavi2024} (see their Table S1), with isotopologue ratios of 70 and 35 for species with one and two carbon atoms respectively, following the carbon fractionation ratio reported by \citet{2009ApJ...693.1360W}.

To perform an accurate comparison between the slab models and the MIRI spectrum of TWA~27A, the continuum emission has to be subtracted from the spectrum. After removing the emission from the star (see mid panel of Figure~\ref{fig:twa-27-A_modelfit}) we initially tried to guide the continuum between 7 and 20~$\mu$m identifying by eye line-free regions of the spectrum to then use a cubic spline interpolation to determine the continuum level. However, based on the residuals from the initial fitting of the molecular features, the presence of molecular pseudo-continuum was suspected. Given the absence of a silicate dust feature at $\sim$10~$\mu$m, and the difficulty of distinguishing the molecular pseudo-continuum from a multi-blackbody fit, we follow the analysis procedure outlined by \citet{Arabhavi2024}, who similarly analysed the spectrum of ISO-ChaI 147 dominated by molecular pseudo-continuum without dust features. Accordingly, we define the continuum as a straight line between 8.46 and 18~$\mu$m, adjusting its slope slightly after the first fit to match the residual flux at 8.46~$\mu$m.

To find the best slab model fit we used the \texttt{prodimopy} (v2.1.4)\footnote{Available at \url{https://gitlab.astro.rug.nl/prodimo/prodimopy}} Python package to perform reduced $\chi^2$ fits of the convolved and resampled slab models to the continuum subtracted spectrum of TWA~27A. In this procedure, the only free parameters are the column density (N), the temperature (T), and the equivalent radius (R) of the emitting area ($\pi$R$^2$). The column density and temperature determine the relative flux levels of the emission lines, while the emitting radius is a scaling factor to match the absolute flux level of the observation. The $\chi^2$ fits are done iteratively, one molecule at a time. For each molecule, we select windows \citep[see ][]{Arabhavi2024} for the $\chi^2$ that are the least affected by other species but which still contain enough spectral features to successfully constrain the fit. In this way, we minimize the contribution of other species. In the case of C$_2$H$_4$ and C$_2$H$_2$ there is a dichotomy between a model with large column density which reproduces well the molecular pseudo-continuum and another model with lower column density which reproduces the peaks of the $Q$-branches. In these cases we treated the affected molecule as if it was two different species fitted together but we limit the parameter space of the column density of one model to larger column densities and the other model to lower column densities. We assume that the temperature is the same for both models. In the case of C$_3$H$_4$ and C$_4$H$_2$ the emission of both species overlap. In this case we fitted both molecules simultaneously leaving the column density and temperature as free parameters.

The order of the fitting process is: C$_2$H$_4$ (2 models of higher and lower column densities), C$_2$H$_2$ + $^{13}$CCH$_2$, (2 models of higher and lower column densities), C$_6$H$_6$, CO$_2$ + $^{13}$CO$_2$, C$_3$H$_4$ + C$_4$H$_2$ (fitted together), C$_2$H$_6$, HC$_3$N, HCN, and CH$_4$. CH$_4$ also presents some molecular pseudo-continuum but we do not need a second model to fit it. We followed the criteria in Morales-Calder\'on et al. (submitted) to determine if a detection is real. If at least two features are detected, the main one is stronger than the 3$\sigma$ uncertainty and the same molecule is detected if we change the order of the fitting we consider it a firm detection. All the detections presented here are firm. While line overlap between lines of the same molecule are accounted for in our slab models, the line overlap between different molecules are not taken into account in our fitting procedure. Thus, the errors in the fit are passed to the next molecule and we cannot estimate the properties or reliable upper limits for molecules emitting in the same spectral region. This is the case of C$_2$H$_6$, HC$_3$N, HCN which lie on top of the pseudo-continuum of other molecules. While we can consider these as firm detections, we cannot provide accurate parameters. 

\begin{figure*}
   \centering
    \includegraphics[width=\hsize]{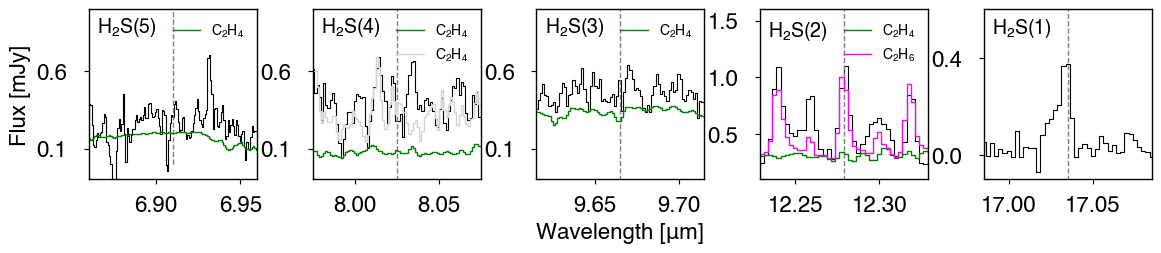}
      \caption{Zooms-in at the positions of the H$_2$ lines in the TWA 27 A  MIRI spectrum. The green, grey and pink lines represent the slab models for C$_2$H$_4$, CH$_4$, and C$_2$H$_6$, respectively. Models are stacked on top of each other. }
         \label{Fig:H2}
   \end{figure*}

The best-fit model is presented in Figure~\ref{fig:twa-27-A_molfit} together with the continuum-subtracted spectrum. The best-fit model parameters are given in Table~\ref{Tab:molecules}, together with the windows in which the $\chi^2$ fits have been evaluated. Notably we do not detect H$_2$O or CO emission from the disk. However, this does not necessarily mean that these species are absent in the spectrum. Stellar photospheric absorption dominates the spectrum at short wavelengths ($<$7\,$\mu$m) where the rovibrational bands of CO and H$_2$O emit, as shown in Fig.\,\ref{fig:twa-27-A_modelfit}. Further, \citet{2025ApJ...984L..62A} show that bright hydrocarbon emission can outshine emission from large column densities of water. In Figure~\ref{fig:twa-27-A_molfit} we include CH$_3$. Because of very few spectroscopic data for this molecule are available we could not fit it in the same way and thus we only show a model for comparison.
In addition we have searched for H$_2$ pure rotational lines in the MIRI-MRS spectrum (see Fig.~\ref{Fig:H2}). While we find a peak at $\sim$17.035\,$\mu$m that corresponds to the H$_2$ S(1) line position, we do not detect any other H$_2$ lines.  For example, H$_2$ S(2) falls in the same spectral area as the C$_2$H$_6$ emission as can be seen by the pink line in  Fig.~\ref{Fig:H2} which represents the C$_2$H$_6$ slab model. The slab model of C$_2$H$_6$ is, in turn, not well constrained because it lies on top of the pseudo-continuum formed by the emission of C$_2$H$_4$ (green line in Fig.\ref{Fig:H2}). An in depth modelling of this emission is needed in order to quantify the presence of the emission of H$_2$ at that wavelength.

%-------------------------------------------------------------

% _____________________________________________________________

% _____________________________________________________________

\begin{table}[h!]
\caption{Temperatures, column densities, emitting radii and the window where the $\chi^2$ was evaluated for each of the detected molecules obtained
from the slab model fits.}             % title of Table
\label{Tab:molecules}      % is used to refer this table in the text
\centering                          % used for centering table
\begin{tabular}{l c c c c}        % centered columns (4 columns)
\hline\hline                 % inserts double horizontal lines
Species & $T[K]$ & N[cm$^{-2}$] & R[au] & Window[\mum]\\    % table heading
\hline                        % inserts single horizontal line
C$_2$H$_4$ & 375 & 6.81$\times$10$^{23}$ & 0.005 & 9.5 - 11.5 \\      % inserting body of the table

C$_2$H$_4$ & 375 & 2.15$\times$10$^{18}$ & 0.009 & 9.5 - 11.5 \\      % inserting body of the table

C$_2$H$_2$ + $^{13}$CCH$_2$ & 300 & 1.00$\times$10$^{21}$ & 0.010  & 12.6 - 13.71\\      % inserting body of the table

C$_2$H$_2$ + $^{13}$CCH$_2$ & 300 & 6.81$\times$10$^{17}$ & 0.025  & 12.6 - 13.71\\      % inserting body of the table

C$_6$H$_6$ & 350 & 2.15$\times$10$^{16}$\tablefootmark{*}    & 0.034  & 14.83 - 14.88\\

CO$_2$ +  $^{13}$CO$_2$ & & & & 14.925 - 15 \\
CO$_2$ +  $^{13}$CO$_2$ & 275 & 4.65$\times$10$^{18}$     & 0.036 & 15.36 - 15.462\\
CO$_2$ +  $^{13}$CO$_2$ & & & & 16.16 - 16.21 \\
C$_3$H$_4$ & 200 & 6.81$\times$10$^{17}$\tablefootmark{*}  & 0.028 & 15.7 - 16.0 \\      % inserting body of the table

C$_4$H$_2$& 250 & 4.65$\times$10$^{16}$\tablefootmark{*} &  0.027  & 15.7 - 16.0\\

C$_2$H$_6$ & \multicolumn{3}{c}{Detected} & 11.9 - 12.4\\ 

HC$_3$N  & \multicolumn{3}{c}{Detected} & 15.0 - 15.2\\

HCN &   \multicolumn{3}{c}{Detected} &13.85 - 14.325 \\

CH$_4$ & 375 & 6.81$\times$10$^{20}$ & 0.010 &   7.0 - 9.0 \\   % inserting body of the table

CH$_3$ & \multicolumn{3}{c}{Detected} & \\ 

\hline                                   %inserts single line
\end{tabular}
\tablefoot{
%\tablefoottext{a}{Tentative detection.} \\
\tablefoottext{*}{Parameter not well constrained.} 
}
\end{table}

% \subsection{}
\section{Discussion}\label{sec:discussion}
\subsection{TWA~27A atmosphere}
As shown in Fig.\,\ref{fig:twa-27-A_modelfit}, absorption features from the TWA~27A atmosphere dominates the MIRI spectrum up to $\sim$6.5\,$\mu$m, beyond which the molecular emission from the disk starts to dominate. As shown by \citet{Arabhavi2024}, \citet{Kanwar2024}, and \citet{2025ApJ...984L..62A}, the short wavelength end of the MIRI spectrum shows the CO and water absorption features from the atmosphere, although there is an overall infrared excess emission at these wavelengths.

\subsection{TWA~27A disk}
The inner disk gas of TWA 27 A is hydrocarbon-rich, all detected molecules contain carbon, while CO$_2$ (and $^{13}$CO$_2$) is the only oxygen-bearing molecule. The diversity of hydrocarbons indicates that the C/O ratio is larger than unity. Assuming that the emitting radii indicate the physical location of the molecular emission, the emission is confined to the disk region that is closer than 0.1\,au to the star.

The chemical composition of the TWA 27 A disk emitting layer and the estimated column densities and temperatures are very similar to those found in the younger Chamaeleon-I source ISO-ChaI\,147 \citep{Arabhavi2024} and in the much older source WISE\,J044634.16-262756.1 in the Columba association \citep{2024arXiv241205535L}. The shape of the 10\,$\mu$m region is also similar in these sources. The gas composition of TWA 27 A probed by MIRI is more diverse compared to J160532 \citep{Tabone2023} and Sz28 \citep{Kanwar2024}.
 
% \ama{Add a zoom-in on the residuals between 4-7 microns, either in main text or appendix, to show possible CO or H2O emission from the disk}.

Pebble transport models such as \citet{2023A&A...677L...7M}, suggest that the disks around very low-mass stars ($<$0.3\,$M_{\odot}$) become carbon-rich (C/O$>$1) within 3\,Myr, and then remain so for the rest of the disk lifetime. The carbon-rich spectra of the $\sim$10\,Myr old TWA~27A disk suggests that it would be well past the carbon enrichment timescale. Furthermore, ALMA observations of TWA 27 A indicate a dust mass of $\sim$0.1\,$M_{\oplus}$ \citep{Ricci2017}, which is some of the lowest disk dust masses \citep{2016ApJ...831..125P}, which supports the rapid transport of material inward suggested by pebble transport models \citep{2023A&A...677L...7M,2020A&A...638A..88L}. For more detailed comparison of TWA 27 A with a larger sample refer Arabhavi et al. (subm.) and \citet{2025ApJ...984L..62A}.

\subsection{Evidence of a disk around TWA~27b}
% \begin{figure}[h]
%     \centering
%     \includegraphics[width=0.5\textwidth]{figures/06_twa-27_b_empirical_disk.png}
%     \caption{Top: Comparison of TWA~27b with other low gravity gas giant and brown dwarfs scaled to the distance of TWA~27. The 15 \mum flux is consistently higher than what would be explained by just the photosphere by at least 3 $\sigma$. Bottom: Comparison of TWA~27b with the most similar spectral type L6 brown dwarfs 2M~2148, with the addition of the predicted CPD emission of a 1300~K object from \cite{Sun2024}. The overall spectrum of 2M~2148 plus the transitional disk model emission fits the observed data the best.}
%     \label{fig:twa-27-b_empirical_comparison}
% \end{figure}
As discovered by \cite{Luhman2023_twa27} and analysed by \cite{Marleau2024}, emission lines due to accretion of gas point to the existence of a CPD around TWA~27b. With low accretion rate and very low mass limits for the dust in the disk with ALMA \citep{Ricci2017}, one would expect a late-stage disk around the companion, with its emission shifted towards the longer wavelengths. 

Due to the fact that the self-consistent grid models do not reproduce the silicate cloud feature in MIRI, the IR excess at 15 \mum is difficult to interpret from the results of Section~\ref{sec:analysis_b}. The excess could for example be a result of discrepancy in the PT structure between data and models. We attempted to fit the atmosphere together with the disk blackbody contribution by using the spectra up to 6.5 \mum (in order to avoid the silicate feature) and adding the 15 \mum flux, similar to the analysis of TWA~27A. The fit did not converge to a reasonable solution, adding a blackbody component of 1000~K and emitting area of 1~$\RJ$ while still not fitting the 15 \mum flux. This could be interpreted as the optimisation trying to deal with the muted features of the atmosphere by combining a clear atmosphere (self-consistent grid model) with a cloudy featureless atmosphere (blackbody component).

\begin{figure}[h]
    \centering
    \includegraphics[width=0.5\textwidth]{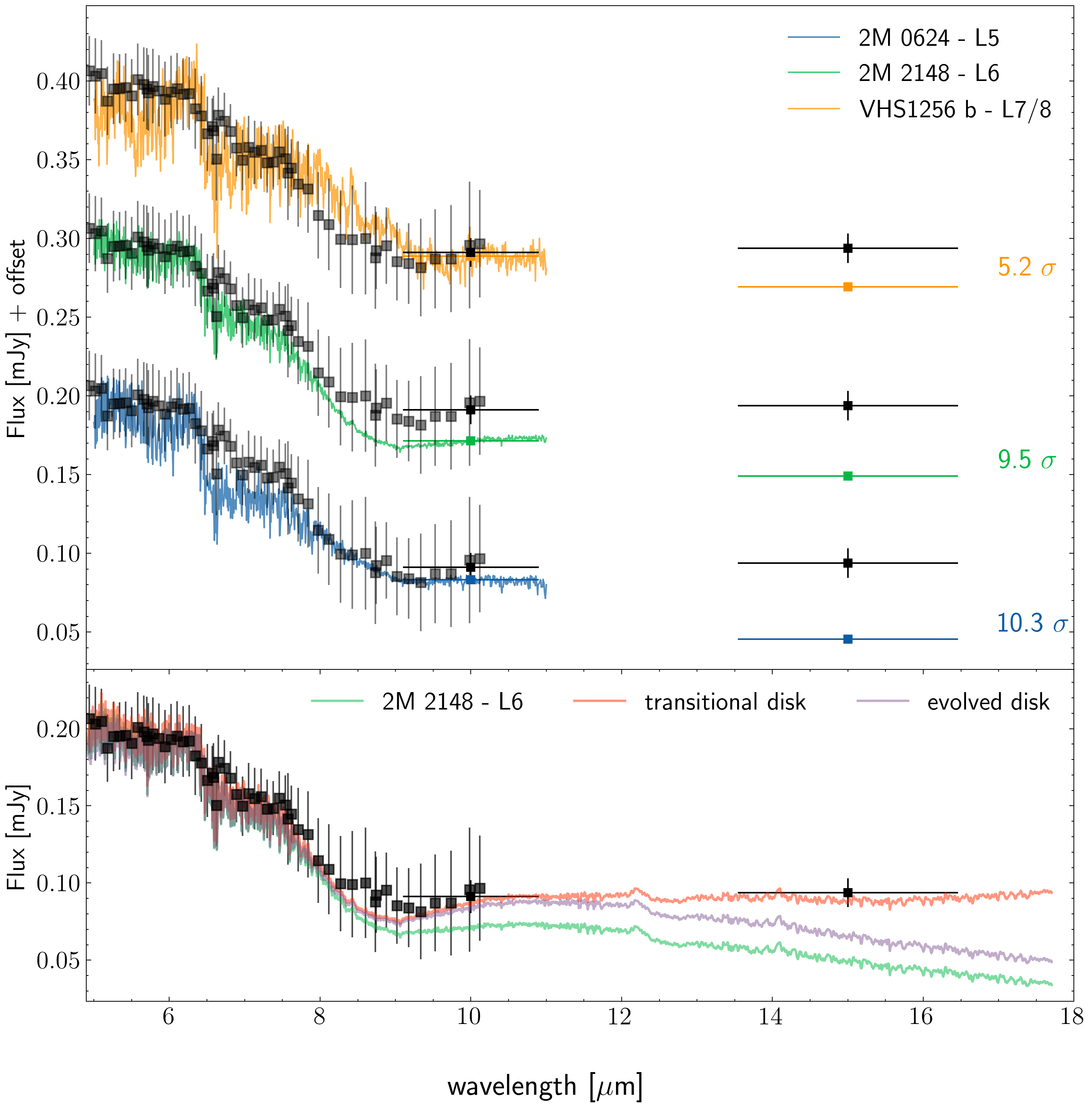}
    \caption{Top: Comparison of TWA~27b with other low gravity gas giant and brown dwarfs scaled to the distance of TWA~27. The 15 \mum flux is consistently higher than what would be explained by just the photosphere by at least 3 $\sigma$. Bottom: Comparison of TWA~27b with the most similar spectral type L6 brown dwarfs 2M~2148, with the addition of the predicted CPD emission of a 1300~K object from \cite{Sun2024}. The overall spectrum of 2M~2148 plus the transitional disk model emission fits the observed data the best.}
    \label{fig:twa-27-b_empirical_comparison}
\end{figure}

We proceeded to compare the TWA~27b SED with publicly available MIRI/MRS spectra of mid to late L-dwarfs. Similar to \cite{Luhman2023_twa27}, we compared to the PMC VHS~1256~b from the ERS program 1386 \citep{Miles2023} that has a spectral type of L7/8,
%PSO~318 from PID~1275 \citep[PI: Lagage;][in prep.]{Molliere2025_pso}, 
and two more isolated brown dwarfs of similar spectral type from PID~2288 \citep[PI: Lothringer;][in prep.]{Valenti2024_bds}. 
%, while PSO~318 is a L7 isolated planetary mass object \todo{ref}. 
The brown dwarf 2M~0624-4521 has a spectral type L5 \citep{Parker2013, Manjavacas2016}, while 2M~2148+4003 is an L6 dwarf \citep{Looper20082m2148, Allers2013}. We scaled the flux of these empirical spectra only by the distance ratio (D$_{\text{object}}^2$/D$_{\text{TWA~27}}^2$) and show the comparison in Figure~\ref{fig:twa-27-b_empirical_comparison}. The spectra shown up to 11 \mum have been binned slightly to improve visibility, and offset by 0.1 mJy. The 15 \mum flux was calculated using \texttt{species}, integrating the MRS spectrum over the F1500W filter, including propagation of the uncertainty which is not visible in the plot.

We observe that compared to all the objects the 15 \mum flux of TWA~27b shows an IR excess with a significance of $\sim$5-10~$\sigma$ (Figure~\ref{fig:twa-27-b_empirical_comparison} top panel). Interestingly, TWA~27b matches very well its NIR spectral type L6 \citep{Manjavacas2024} counterpart 2M~2148 in the 5-7 \mum range but then starts diverging and showing excess emission. The L5 object looks a bit too blue, while the L7/8 planetary mass objects seems redder. We note that due to the large uncertainty of the MRS spectrum, these differences all fall within 1-2~$\sigma$. 

In order to assess whether the excess seen at 15 \mum would be physically plausible we compare to predictions of the disk flux around such a low mass object \cite{Zhu2015, Chen2022cpds, Sun2024}. We used models from \cite{Sun2024} that predict the expected flux of the disk for various evolution stages with a central source corresponding to a 1300 K gas giant. In the bottom panel of Figure~\ref{fig:twa-27-b_empirical_comparison} we show the comparison of TWA~27b with 2M~2148, and the flux of 2M~2148 plus the excess emission due to a transitional and evolved CPD from \cite{Sun2024}. The transitional disk model shows the best agreement with our data at 15 \mum, and both disk models match the 7-11 \mum silicate clouds region better. Given the auxiliary evidence of accretion tracers, the MIRI data support the existence of a CPD around TWA~27b that provides the mass reservoir for the gas accreted onto the companion. Due to the complex nature of L/T transition type atmospheres, a potential solution to the 15 \mum IR-excess could be found by a free atmospheric retrieval that accounts for both the silicate cloud  composition and potential thermal structure of the atmosphere.

\subsection{Atmosphere and clouds of TWA~27b}
The self-consistent atmospheric modelling presented in Section~\ref{sec:analysis_b} and Figure~\ref{fig:twa-27-b_modelfit} showcase the complexity of consistently fitting the atmospheres of young, low gravity L/T transition gas giants. The small uncertainties of spectra obtained with JWST, wide wavelength coverage, and complex physical processes governing the atmospheres do not allow one model to fit all regions equally well. As also explored in \cite{Petrus2024}, different models yield different answers even to fundamental parameters such as temperature and gravity. In this work we tested three grids that have been optimised for such young gas giants and overall \texttt{ExoREM} appears to fit best (reduced $\chi^2 \sim9.8$, but all give similar estimates for T$_{\text{eff}}\sim$1400~K, log(\textit{g})$\sim$4, and R$_{\text{p}}\sim$1.2~$\RJ$. 

The inclusion of extinction in the model seems to improve the fit and yields parameters closer to the expectations by evolutionary models \citep{Marley2021}. Adding extinction to the spectra of young L/T objects has been successful \citep[eg., ][]{Stolker2021}, but a definitive physical explanation of the source of dust that causes the extinction has not been determined. One option would be extinction caused by the CPD, but this would only work for specific inclinations, and only for young objects like TWA~27b that still retain a disk. Another option is extinction due to small-sized dust in the upper atmosphere that causes the deep silicate feature seen in MIRI and is small enough to absorb NIR radiation from the deeper atmosphere layers. Such small suspended dust could be a distinct feature of the low gravity and youth of TWA~27b, not yet included in the self-consistent atmosphere models.

From the MIRI observations presented in this work, the silicate  between 7.5 and 10 \mum is the most intriguing new feature of its atmosphere. As first observed for planetary mass objects in VHS~1256~b, the silicate clouds is a key component of the atmosphere, affecting the full spectrum but can only be constrained at these wavelengths. This is due to the fact that the silicate cloud grain size distribution and composition of the dust can be estimated using the shape and depth of the feature. From Figure~\ref{fig:twa-27-b_empirical_comparison}, and excluding the potential contribution of the CPD, it is obvious that the silicate cloud of TWA~27b has a distinct shape from the planetary mass object VHS~1256~b, better matching the L6 brown dwarf 2M~2148. The silicate feature shifted towards the blue could potentially be an indication of contribution of SiO$_{\text{2}}$ grains \citep{Henning2010, Burningham2021}. Both \cite{Suarez2023} and \cite{Petrus2024} explore the shape and depth of the silicate s in brown dwarfs and PMCs using a spectral index as a function of spectral type, which could potentially be applied to TWA~27b after removing the emission by the potential CPD.

% \subsection{Formation}
% \subsection{Evolutionary models}

\section{Summary and Conclusion}\label{sec:conclusion}

In this work we present mid-IR observations of the first directly detected gas giant, 20 years after its discovery. Both components of the system show novel features in their spectra, summarised here:
\begin{itemize}
    \item The spectrum of the primary M8 brown dwarf TWA~27A is dominated by emission from its disk. At least 11 organic molecules are detected in emission, with C$_{\text{2}}$H$_{\text{2}}$ being the most prominent feature in the spectrum.
    \item We do not detect H$_{\text{2}}$O in the disk of A, with the only oxygen bearing species being CO$_{\text{2}}$, and its isotopologue $^{\text{13}}$CO$_{\text{2}}$. This finding follows the trend of very carbon rich disks around low mass stars and brown dwarfs seen with JWST/MIRI.
    \item Using high contrast imaging methods applied to the MRS, we extracted the spectrum of the planetary mass companion TWA~27b. The spectrum is consistent with its predicted spectral type, however fitting the full 1-15 \mum SED with self-consistent atmosphere models remains challenging. 
    \item One of the key features discovered in the MIRI spectrum of TWA~27b is a broadband absorption feature attributed to silicate clouds condensing in its atmosphere, hinting to a dusty atmosphere. 
    \item We find that the flux at the longest wavelengths, and especially the very precise 15 \mum flux from the MIRI Imager, shows excess emission compared to the expected tail of the blackbody. This implies the existence of a CPD, with first comparisons to models indicating a transitional disk, consistent with the age of the system.
\end{itemize}

The spectra of the TWA~27 system present a challenge in modelling complexity that JWST has started to reveal. With the high quality of data and wide wavelength coverage for objects previously not accessible, the missing physics in models can be highlighted (see for example \citep{Kuhnle2025, Zhang2025}). 

For TWA~27A, the addition of NIRSpec to the rich MIRI spectrum, and even ALMA detection of the disk at milli-meter wavelengths, offers the opportunity to model the disk at much higher complexity by fitting simultaneously the brown dwarf photosphere, the disk continuum (especially in the 3-6 \mum region where the atmosphere and disk emission are comparable), and the disk molecular emission. This will place much tighter constraints on the disk geometry, thermal structure, dust properties, and molecular column densities, potentially breaking degeneracies in the models. 

In the case of TWA~27b, it is the first time that such a low mass and young PMC shows evidence for silicate clouds and a CPD. As shown seen in recent modelling of young low gravity gas giants, often including clouds and disequilibrium chemistry, is a challenge both for self-consistent models and free retrievals  \citep[eg.,][]{Petrus2024}. In order to properly constrain the properties of the atmosphere and disk, a joint modelling of the emission from the atmosphere including silicate cloud species, and the contribution of the disk must be undertaken. Additional challenges such as accounting for the different S/N over the wavelength range need to be better understood \citep[eg., ][]{Barrado2023}, in order for the model fit all wavelength regions (eg., matching the silicate absorption). This fully covered by JWST exoplanetary system will be a benchmark for studying planet formation, at the intersection of protoplanetary disk and atmosphere modelling.

% \begin{itemize}
%     \item system has detected outflow
%     \item 2 papers on accretion ratios
%     \item add shapes of silicate species
%     \item Sun 2024 CPDs

% \end{itemize}
\begin{acknowledgements}
\input{acknowledgements}
\end{acknowledgements}
\bibliography{twarefs} % bibliography data in aanda.bib
\bibliographystyle{aa} 

\end{document}

%% file: authors.tex
\author{
P. Patapis\orcidlink{0000-0001-8718-3732}\inst{\ref{eth}}
\and
M. Morales-Calder\'on \orcidlink{0000-0001-9526-9499}\inst{\ref{cab}}
\and
A. M. Arabhavi \orcidlink{0000-0001-8407-4020}\inst{\ref{kapteyn}}
\and 
H. K\"{u}hnle \orcidlink{0009-0001-2738-2489}\inst{\ref{eth}}
\and 
D. Gasman \orcidlink{}\inst{\ref{kuleuven}}
\and
G. Cugno \orcidlink{0000-0001-7255-3251}\inst{\ref{michigan}, \ref{uzh}}
\and
P. Molli\`{e}re \orcidlink{0000-0003-4096-7067}\inst{\ref{mpia}}
\and
E. Matthews \orcidlink{0000-0003-0593-1560}\inst{\ref{mpia}}
\and
M. M\^alin\orcidlink{0000-0002-2918-8479}\inst{\ref{jhu}, \ref{stsci}}
\and
N. Whiteford\orcidlink{0000-0001-8818-1544}\inst{\ref{museum}}
\and
P.-O. Lagage\orcidlink{}\inst{\ref{saclay}}
\and
R. Waters\orcidlink{}\inst{\ref{radboud}, \ref{hfml}, \ref{sron}}
\and
M. Guedel\orcidlink{0000-0001-9818-0588}\inst{\ref{vienna},\ref{eth}}
\and
Th. Henning\orcidlink{0000-0002-1493-300X}\inst{\ref{mpia}}
\and
B. Vandenbussche\orcidlink{0000-0002-1368-3109}\inst{\ref{kuleuven}}
\and
% Full members and guest authors
O. Absil\orcidlink{0000-0002-4006-6237}\inst{\ref{star}}
\and
I. Argyriou\orcidlink{0000-0003-2820-1077}\inst{\ref{kuleuven}}
\and
D. Barrado\orcidlink{0000-0002-5971-9242}\inst{\ref{cab}}
\and
P. Baudoz \orcidlink{0000-0002-2711-7116}\inst{\ref{lesia}}
\and
A. Boccaletti\orcidlink{0000-0001-9353-2724}\inst{\ref{lesia}}
\and
J. Bouwman\orcidlink{0000-0003-4757-2500}\inst{\ref{mpia}}
\and
C. Cossou\orcidlink{0000-0001-5350-4796}\inst{\ref{saclay}}
\and
A. Coulais\orcidlink{}\inst{\ref{lerma}}
\and
L. Decin\orcidlink{0000-0002-5342-8612}\inst{\ref{kuleuven}}
\and
R. Gastaud\orcidlink{}\inst{\ref{saclay}}
\and
A. Glasse\orcidlink{0000-0002-2041-2462}\inst{\ref{atc}}
\and
A. M. Glauser\orcidlink{0000-0001-9250-1547}\inst{\ref{eth}}
\and
S. Grant \orcidlink{0000-0002-4022-4899}\inst{\ref{mpie}, \ref{carnegie}}
\and
M. Min\orcidlink{0000-0001-5778-0376}\inst{\ref{sron}}
\and
I. Kamp \orcidlink{0000-0001-7455-5349}\inst{\ref{kapteyn}}
\and
% S. Kendrew\orcidlink{0000-0002-7612-0469}\inst{\ref{esa}}
% \and
%O. Krause\orcidlink{}\inst{\ref{}}
%\and
G. Olofsson\orcidlink{0000-0003-3747-7120}\inst{\ref{stockholm}}
\and
J. Pye\orcidlink{0000-0002-0932-4330}\inst{\ref{leicester}}
\and
D. Rouan\orcidlink{0000-0002-2352-1736}\inst{\ref{lesia}}
\and
P. Royer\orcidlink{0000-0001-9341-2546}\inst{\ref{kuleuven}}
\and
S. Scheithauer\orcidlink{0000-0003-4559-0721}\inst{\ref{mpia}}
\and
X. Sun\orcidlink{0009-0005-0743-7031} \inst{\ref{zhuhai}}
\and
P. Tremblin\orcidlink{0000-0001-6172-3403}\inst{\ref{saclay2}}
\and
%co-PIs
L. Colina\orcidlink{0000-0002-9090-4227}\inst{\ref{cab2}}
\and
T. P. Ray\orcidlink{0000-0002-2110-1068}\inst{\ref{dublin}}
\and
G. Östlin\orcidlink{0000-0002-3005-1349}\inst{\ref{oskar}}
\and
E. F. van Dishoeck\orcidlink{}\inst{\ref{leiden}, \ref{mpie}}
\and
G. Wright\orcidlink{0000-0001-7416-7936}\inst{\ref{atc}}
}

%% file: affiliations.tex
\institute{
[1] ETH Zürich, Institute for Particle Physics and Astrophysics, Wolfgang-Pauli-Str. 27, 8093 Zürich, Switzerland \label{eth}\\
\email{polychronis.patapis@phys.ethz.ch} \and
[2] Centro de Astrobiología (CAB), CSIC-INTA, ESAC Campus, Camino Bajo del Castillo s/n, 28692 Villanueva de la Cañada, Madrid, Spain \label{cab}\and
[3] Kapteyn Astronomical Institute, University of Groningen, P.O. Box 800, 9700 AV Groningen, The Netherlands \label{kapteyn}\and
[4] Institute of Astronomy, KU Leuven, Celestijnenlaan 200D, 3001 Leuven, Belgium \label{kuleuven} \and
[5] Department of Astronomy, University of Michigan, Ann Arbor, MI 48109, USA \label{michigan}\and
[6] Department of Astrophysics, University of Zurich, Winterthurerstrasse 190, 8057 Zurich, Switzerland \label{uzh}\and
[7] Max-Planck-Institut für Astronomie (MPIA), Königstuhl 17, 69117 Heidelberg, Germany \label{mpia}\and
Department of Physics \& Astronomy, Johns Hopkins University, 3400 N. Charles Street, Baltimore, MD 21218, USA\label{jhu}\and
[9] Space Telescope Science Institute, 3700 San Martin Drive, Baltimore, MD 21218, USA \label{stsci}\and
[10] Department of Astrophysics, American Museum of Natural History, New York, NY 10024, USA \label{museum} \and
[8] Université Paris-Saclay, Université Paris Cité, CEA, CNRS, AIM, F-91191 Gif-sur-Yvette, France \label{saclay}\and
[11] Department of Astrophysics/IMAPP, Radboud University, PO Box 9010, 6500 GL Nijmegen, The Netherlands \label{radboud} \and
[12] HFML - FELIX, Radboud University, PO Box 9010, 6500 GL Nijmegen, The Netherlands \label{hfml} \and
[13] SRON Netherlands Institute for Space Research, Niels Bohrweg 4, 2333 CA Leiden, The Netherlands \label{sron}\and
[14] Department of Astrophysics, University of Vienna, Türkenschanzstr. 17, 1180 Vienna, Austria \label{vienna}\and
[19] STAR Institute, Université de Liège, Allée du Six Août 19c, 4000 Liège, Belgium \label{star}\and
[20] LESIA, Observatoire de Paris, Université PSL, CNRS, Sorbonne Université, Université de Paris Cité, 5 place Jules Janssen, 92195 Meudon, France \label{lesia}\and
[22] LERMA, Observatoire de Paris, Université PSL, CNRS, Sorbonne Université, Paris, France \label{lerma}\and
[18] Leiden Observatory, Leiden University, P.O. Box 9513, 2300 RA Leiden, The Netherlands \label{leiden}\and
[23] UK Astronomy Technology Centre, Royal Observatory Edinburgh, Blackford Hill, Edinburgh EH9 3HJ, UK \label{atc}\and
Earth and Planets Laboratory, Carnegie Institution for Science, 5241 Broad Branch Road, NW, Washington, DC 20015, USA \label{carnegie} \and
[32] Max-Planck-Institut für Extraterrestrische Physik, Giessenbach- strasse 1, 85748 Garching, Germany \label{mpie} \and
[24] European Space Agency, Space Telescope Science Institute, Baltimore, MD, USA \label{esa}\and
[26] Department of Astronomy, Stockholm University, AlbaNova University Center, 106 91 Stockholm, Sweden \label{stockholm}\and
[15] School of Physics \& Astronomy, Space Park Leicester, University of Leicester, 92 Corporation Road, Leicester, LE4 5SP, UK \label{leicester}\and
[31] School of Physics and Astronomy, Sun Yat-sen University, Zhuhai 519082, People's Republic of China \label{zhuhai} \and
[27] Université Paris-Saclay, UVSQ, CNRS, CEA, Maison de la Simulation, 91191, Gif-sur-Yvette, France \label{saclay2}\and
[28] Centro de Astrobiología (CAB, CSIC-INTA), Carretera de Ajalvir, 8850 Torrejón de Ardoz, Madrid, Spain \label{cab2}\and
[29] School of Cosmic Physics, Dublin Institute for Advanced Studies, 31 Fitzwilliam Place, Dublin, D02 XF86, Ireland \label{dublin}\and
[30] Department of Astronomy, Oskar Klein Centre, Stockholm University, 106 91 Stockholm, Sweden \label{oskar}\and
}

%% file: acknowledgements.tex
The authors wanted to thank K. Luhman and E. Manjavacas for sharing their NIRSpec spectra early. We thank R. Dong, G. Marleau, V. Christiaens for our fruitful discussions. 
\newline
\newline
   \textbf{Contributions:} PP led the project, performed the data analysis, atmosphere model fitting, writing of the manuscript, and created the figures. MCM and AMA performed the molecular modelling of the disk, and contributed to the writing of the manuscript. HK performed atmospheric modelling of the spectra. GC developed the high contrast imaging pipeline for the MRS. DG and SG provided insights early in the project on the disk molecular emission. XS provided the CPD disk models. PM, EM, MM OA provided critical early feedback on the project. POL led the MIRI Exoplanet WG. All authors read and provided feedback on the manuscript.

\newline
\newline
This research has made use of the NASA Astrophysics Data System and the \texttt{python} packages \texttt{numpy} \citep{harris2020array}, \texttt{scipy} \citep{2020SciPy-NMeth}, \texttt{matplotlib} \citep{Hunter:2007} and \texttt{astropy} \citep{astropy:2013, astropy:2018}.
\newline
This work is based on observations made with the NASA/ESA/CSA James Webb Space Telescope. The data were obtained from the Mikulski Archive for Space Telescopes at the Space Telescope Science Institute, which is operated by the Association of Universities for Research in Astronomy, Inc., under NASA contract NAS 5-03127 for JWST. These observations are associated with programs 1029, 1270, 1050, 1386,  1524, 1536, 1538, 2288, 4487. 
MIRI draws on the scientific and technical expertise of the following organisations: Ames Research Center, USA; Airbus Defence and Space, UK; CEA-Irfu, Saclay, France; Centre Spatial de Liége, Belgium; Consejo Superior de Investigaciones Científicas, Spain; Carl Zeiss Optronics, Germany; Chalmers University of Technology, Sweden; Danish Space Research Institute, Denmark; Dublin Institute for Advanced Studies, Ireland; European Space Agency, Netherlands; ETCA, Belgium; ETH Zurich, Switzerland; Goddard Space Flight Center, USA; Institute d'Astrophysique Spatiale, France; Instituto Nacional de Técnica Aeroespacial, Spain; Institute for Astronomy, Edinburgh, UK; Jet Propulsion Laboratory, USA; Laboratoire d'Astrophysique de Marseille (LAM), France; Leiden University, Netherlands; Lockheed Advanced Technology Center (USA); NOVA Opt-IR group at Dwingeloo, Netherlands; Northrop Grumman, USA; Max-Planck Institut für Astronomie (MPIA), Heidelberg, Germany; Laboratoire d’Etudes Spatiales et d'Instrumentation en Astrophysique (LESIA), France; Paul Scherrer Institut, Switzerland; Raytheon Vision Systems, USA; RUAG Aerospace, Switzerland; Rutherford Appleton Laboratory (RAL Space), UK; Space Telescope Science Institute, USA; Toegepast- Natuurwetenschappelijk Onderzoek (TNO-TPD), Netherlands; UK Astronomy Technology Centre, UK; University College London, UK; University of Amsterdam, Netherlands; University of Arizona, USA; University of Bern, Switzerland; University of Cardiff, UK; University of Cologne, Germany; University of Ghent; University of Groningen, Netherlands; University of Leicester, UK; University of Leuven, Belgium; University of Stockholm, Sweden; Utah State University, USA. A portion of this work was carried out at the Jet Propulsion Laboratory, California Institute of Technology, under a contract with the National Aeronautics and Space Administration.

PP thanks the Swiss National Science Foundation (SNSF) for financial support under grant number 200020\_200399.
MMC and DB are supported by grants PID2019-107061GB-C61 and PID2023-150468NB-I00  Spanish MCIN/AEI/10.13039/501100011033 by the Spain Ministry of Science and Innovation/State Agency of Research MCIN/AEI/ 10.13039/501100011033 and by “ERDF A way of making Europe”.
I.K., A.M.A., and E.v.D. acknowledge support from grant TOP-1 614.001.751 from the Dutch Research Council (NWO).
P.-O.L., A.B., C.C., A.C., and R.G. acknowledge funding support from CNES.
% P.-O.L., A.B., C.C., A.C., R.G., D.R., and P.T. acknowledge funding support from CNES.
JPP acknowledges financial support from the UK Science and Technology Facilities Council, and the UK Space Agency.
This project has received funding from the European Union’s Horizon 2020 research and innovation programme under the Marie Sklodowska-Curie grant agreement no. 860470.
% BV thanks the European Space Agency (ESA) and the Belgian Federal Science Policy Office (BELSPO) for their support in the framework of the PRODEX Programme.
BV, OA, IA, and PR thank the European Space Agency (ESA) and the Belgian Federal Science Policy Office (BELSPO) for their support in the framework of the PRODEX Programme.
OA is a Senior Research Associate of the Fonds de la Recherche Scientifique – FNRS.% OA thanks the European Space Agency (ESA) and the Belgian Federal Science Policy Office (BELSPO) for their support in the framework of the PRODEX Programme.
% IA thanks the European Space Agency (ESA) and the Belgian Federal Science Policy Office (BELSPO) for their support in the framework of the PRODEX Programme.
% there is probably a common acknowledgment to CNES for all french co authors.
% C.C. acknowledges funding support from CNES.
LD acknowledges funding from the KU Leuven Interdisciplinary Grant (IDN/19/028), the European Union H2020-MSCA-ITN-2019 under Grant no. 860470 (CHAMELEON) and the FWO research grant G086217N.
MPIA acknowledges support from the Federal Ministry of Economy (BMWi) through the German Space Agency (DLR).
GO acknowledges support from the Swedish National Space Board and the Knut and Alice Wallenberg Foundation.
% PR thanks the European Space Agency (ESA) and the Belgian Federal Science Policy Office (BELSPO) for their support in the framework of the PRODEX Programme.
PT acknowledges support by the European Research Council under Grant Agreement ATMO 757858.
LC acknowledges support by grant PIB2021-127718NB-100 from the Spanish Ministry of Science and Innovation/State Agency of Research MCIN/AEI/10.13039/501100011033.
The Cosmic Dawn Center (DAWN) is funded by the Danish National Research Foundation under grant No. 140. TRG is grateful for support from the Carlsberg Foundation via grant No. CF20-0534.
TPR acknowledges support from the ERC through grant no.\ 743029 EASY.
Support from SNSA is acknowledged.
EvD acknowledges support from A-ERC grant 101019751 MOLDISK.
GC thanks the Swiss National Science Foundation for financial support under grant numbers P500PT\_206785 and P5R5PT\_225479. For the purpose of open access, the authors have applied a Creative Commons Attribution (CC BY) licence to the Author Accepted Manuscript version arising from this submission.